\documentclass{aastex}
\usepackage[update,prepend]{epstopdf}
\usepackage{spr-astr-addons}
\usepackage{url}
\usepackage{breakcites}
\usepackage{amsmath}
\usepackage{amsfonts}
\usepackage[breaklinks=true]{hyperref}
\usepackage{graphics}

\begin{document}

\title{Interacting Tsallis holographic dark energy in higher dimensional Cosmology}

\author{A. Saha\altaffilmark{1}}
\altaffiltext{1}{
Physics Department, Jalpaiguri Govornment Engineering College, Jalpaiguri, West Bengal, India 73510 \\
arindamjal@gmail.com}
\author{S. Ghose\altaffilmark{2}}
\altaffiltext{2}{
HECRC, University of North Bengal, Rajarammohunpur, Darjeeling, West Bengal, India 734013 \\dr.souvikghose@gmail.com
}


\begin{abstract}

  We discuss Tsallis holographic dark energy (THDE) model in higher dimension. An interacting dark energy model is proposed with Generalized Chaplygin Gas (GCG) in the framework of Compact Kaluza-Klein gravity. It is shown that a stable configuration can be found in the present epoch which is also compatible with the observed value of the density parameter. It is noted that dark energy (DE) might have evolved from a phantom phase in the past. Nature of dark energy  is found to depend on the coupling parameter of the interaction. Classical stability consideration is also found to put an upper bound on the model parameter.

\end{abstract}


\section{Introduction}
Present observations suggest that our universe is spatially flat [\cite{spergel2003,komatsu2011}] and accelerating [\cite{riess,perlm,perlm2,perlm3}]. There are two common approaches to explain the acceleration of universe.  One approach is to modifiy general relativity such as in $f(R)$, $f(G)$ gravity theories and in scalar tensor theories [\cite{carroll,uddin,linderfg,samipaddy,nojde1,cappode,liucamp}]. The other approach is to consider some form of energy which is capable of generating acceleration. These exotic forms of energy are commonly known as dark energy. The simplest dark energy model is to consider a cosmological constant. However, there is a huge mismatch between theoretically calculated and observationally suggested value of the cosmological constant.  Despite sustained efforts [\cite{nojirieos,bambaeos}] the exact nature of dark energy is still unknown. Consequently many dynamical dark energy models are considered in literature [\cite{ratra1988,gibb,armendariz2001,wetterich2017,escamilla2013}] .  Generalized Chaplygin Gas (GCG hereafter) is one such candidate [\cite{bentosen,cald,mebarki2019,khadekar2019}]. The equation of state for GCG is given as:
\begin{equation}
\label{gcgeos}
p=-\frac{A}{\rho^{\alpha}},
\end{equation}
where $A$ is a positive constant and $0\leq \alpha \leq 1$. At high energy GCG behaves as pressure-less dust but at low energy it behaves as dark energy. GCG can be discussed as a complex scalar field theory originating from Born-Infeld action. Apart from dark energy observations also suggest existence of dark matter and GCG is a good candidate for unified dark energy and dark matter models [\cite{kamenshchik2001,bilic2002,thakur}]. 
Recently, holographic principle has emerged as an essential proposal in quantum gravity. When applied to vacuum, holographic principle leads to Holographic Dark Energy (HDE hereafter) models [\cite{susskind,cohen,lii}] . Holographic principle largely states that \textit{number of degrees of freedom of any physical system should scale with the boundary surface and not with the volume}. HDE models have been widely discussed in literature [\cite{granda2008,gao2009,del2011,lepe2010,arevalo2014}]. One of the characteristic features of HDE models is the long distance cut off, also known as infrared cut off (IR cut off hereafter). This cut off is not uniquely defined in cosmology. One of the most natural choices for IR cut off  is Hubble horizon. But HDE, as noted by Hsu [\cite{hsu2004}], could not give recent acceleration. Later Zimdahl and Pav\'on [\cite{zimdahl2007}] showed that allowing interaction in the dark energy sector completely solved the problem. For a detailed review of different dark energy models see Ref. (\cite{miao2011}).

Another intriguing problem for scientists is to unify gravity with other gauge interactions. This challenge has motivated the study of higher dimensional theories. In many higher dimensional models  more than four dimension of space-time is suggested for early universe (at very high energy Kaluza [\cite{kaluza}] and Klein [\cite{klein}] proposed a five dimensional theory of general relativity with an aim to unify gravity and electromagnetic interaction. The compact version of the theory (KK here after) proposed a 5D universe where subsequent evolution  lead to  expansion of four dimensions while extra dimensions has shrunk due to compactification. The theory inspired many cosmological models in literature as the curvature of the five dimensional space-time also induces effective properties of matter in four dimension [\cite{leekk,appelkk,darabi,faraz}]. Interacting HDE models have been studied in KK framework recently by Sharif and Khanum [\cite{sharifkk}] to show that generalized second law of thermodynamics holds throughout for any FRW type cosmological model. Their conclusion holds for any HDE model and is independent of dark energy equation of state.

In the present work Tsallis Holographic Dark Energy correspondence of interacting GCG model is made in compact KK framework. In the next section (sec. (\ref{sthde})) the motivation for TDHE is discussed. Relevant field equations of KK cosmology are set up in section (\ref{feqkk}). In section (\ref{int GCG})an interacting GCG model with THDE correspondence is proposed and  the evolution of universe in such a model is elaborated. Stability of the model in present epoch is considered in section (\ref{sec:sqv})and in section (\ref{disc}) the present work is concluded with relevant discussion.
\section {Tsallis Holographic Dark Energy}
\label{sthde}
Core of any HDE model lies in the definition of system boundary and one common way to modify HDE models is to change it [\cite{zhang2005,wang2016,wang2017}]. Additionally application of Tsallis statistics to the system horizon leads to Tsallis Holographic Dark Energy (THDE hereafter) models. Tsallis entropy is non extensive in nature. Here the black hole horizon entropy is modified as $S_{\delta}=\gamma A^{\delta}$, where $\delta$ denotes the non-extensive parameter and $\gamma$ is an unknown constant.  Bekenstein entropy can be obtained as a limiting case from Tsallis entropy. THDE models also show good stability by themselves\cite{majhi,abe,touchette,biro}. Quantum gravity considerations have also confirmed functional behaviour of Tsallis's entropy content of a system which is a power law function of system area [\cite{tsalis}]. THDE has been studied with Hubble horizon as IR cut off. These models give accelerating universe but are unstable. There are many other works in literature which discuss THDE with other IR cutoffs [\cite{sheykhi2018,ghaffari2018,ghaffari2019}]. In the present work we use the apparent horizon as IR cut off which coincides with the Hubble horizon. 
\section{Field Equations in the Framework of Kaluza-Klein Cosmology}
\label{feqkk}
\begin{figure}[hbt!]
\includegraphics[scale=0.5]{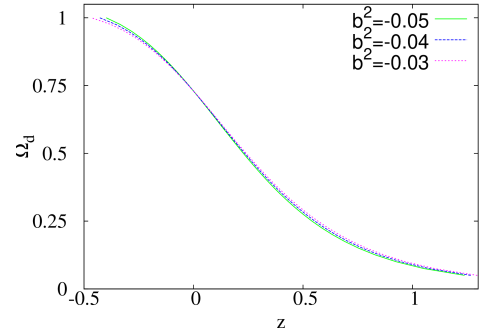}
\centering
\caption{$\Omega_{D}$ vs.$z$ plot for $\delta=2.2$ and different $b^2$}
\label{omz}
\end{figure}
\begin{figure}[hbt!]
 \includegraphics[scale=0.5]{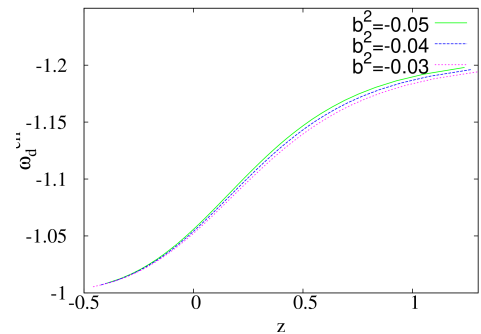}
\centering
 \caption{$\omega^{eff}_{D}$ vs.$z$ plot for  $\delta=2.2$ and different $b^2$}
\label{omeffz}
\end{figure}
Einstein field equation is given by:
\begin{equation}
\label{efeq}
R_{AB}-\frac{1}{2}g_{AB}R=\kappa T_{AB}
\end{equation}
where $A$ and $B$ runs from $0$ to $4$, $R_{AB}$ is the Ricci tensor, $R$ is the Ricci scalar and $T_{AB}$ is the energy-momentum tensor.
The $5$-dimensional space-time metric of Kaluza-Klein (KK) cosmology is (see Ref. \cite{ozel}) :
\begin{eqnarray}
\label{kkmetric}
ds^2 &=& dt^2-a^2(t)\left[\frac{dr^2}{1-kr^2}+r^2(d\theta^2 \right. \nonumber \\
&& \left. + sin^2\theta d\phi^2\right)+\left(1-kr^2)d\psi^2\right],
\end{eqnarray}
where $a(t)$ denotes the scale factor and $k=0,1(-1)$ represents the curvature parameter for flat and closed (open) universe. We consider a cosmological model where KK universe is filled with a perfect fluid.  The Einstein's field equation for the metric given by (\ref{kkmetric}) becomes:
\begin{equation}
\label{kkfeq1}
\rho=6\frac{\dot{a}^2}{a^2}+\frac{k}{a^2},
\end{equation}
\begin{equation}
\label{kkfeq2}
p=-3\frac{\ddot{a}}{a}-3\frac{\dot{a}^2}{a^2}-3\frac{k}{a^2}.
\end{equation}
We consider a flat universe ($k=0$) for simplicity. Eq. (\ref{kkfeq1}) and eq. (\ref{kkfeq2})) then reduce to:
\begin{equation}
\label{kkfeq3}
\rho=6\frac{\dot{a}^2}{a^2},
\end{equation}
\begin{equation}
\label{kkfeq4}
p=-3\frac{\ddot{a}}{a}-3\frac{\dot{a}^2}{a^2}.
\end{equation}
The Hubble parameter is defined as $H=\frac{\dot{a}}{a}$ and $T^{\mu\nu}_{;\nu}=0$ gives the continuity equation:
\begin{equation}
\label{kkcont1}
\dot{\rho}+4H(\rho+p)=0
\end{equation}
Using the equation of state $p=\omega\rho$ in the equation of continuity in five dimension:
\begin{equation}
\label{kkcont2}
\dot{\rho}+4H\rho(1+\omega)=0
\end{equation}
We now consider two types of cosmic fluid. Total energy density is then $\rho=\rho_{D}+\rho_{m}$, where $\rho_{D}$ corresponds to dark energy and $\rho_{m}$ is for matter including Cold Dark Matter (CDM) with $\omega_{m}=0$. For non-interacting fluid, the  conservation equations for $p_{D},\rho_{D}$ and $p_{m},\rho_{m}$ are separately satisfied. For interacting dark energy models:
\begin{equation}
\label{kkcontm}
\dot{\rho_m}+4H\rho_{m}(1+\omega_m)=Q
\end{equation}
\begin{equation}
\label{kkcontl}
\dot{\rho_{D}}+4H\rho_{D}(1+\omega_{D})=-Q
\end{equation}
where $Q$ gives the interaction between dark energy and dark matter.
\section{THDE correspondence of GCG model and evolution of universe}
\label{int GCG}
\begin{figure}
\centering
 \includegraphics[scale=0.5]{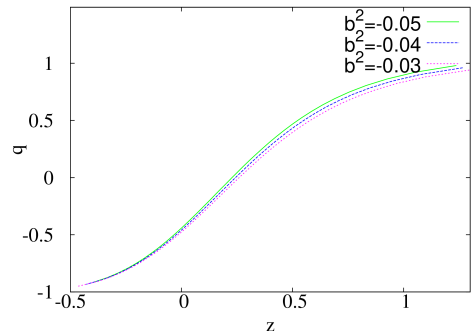}
 \caption{$q$ vs. $z$ plot for  $\delta=2.2$ and different $b^2$}
\label{qz}
\end{figure}

\begin{figure}
\centering
 \includegraphics[scale=0.5]{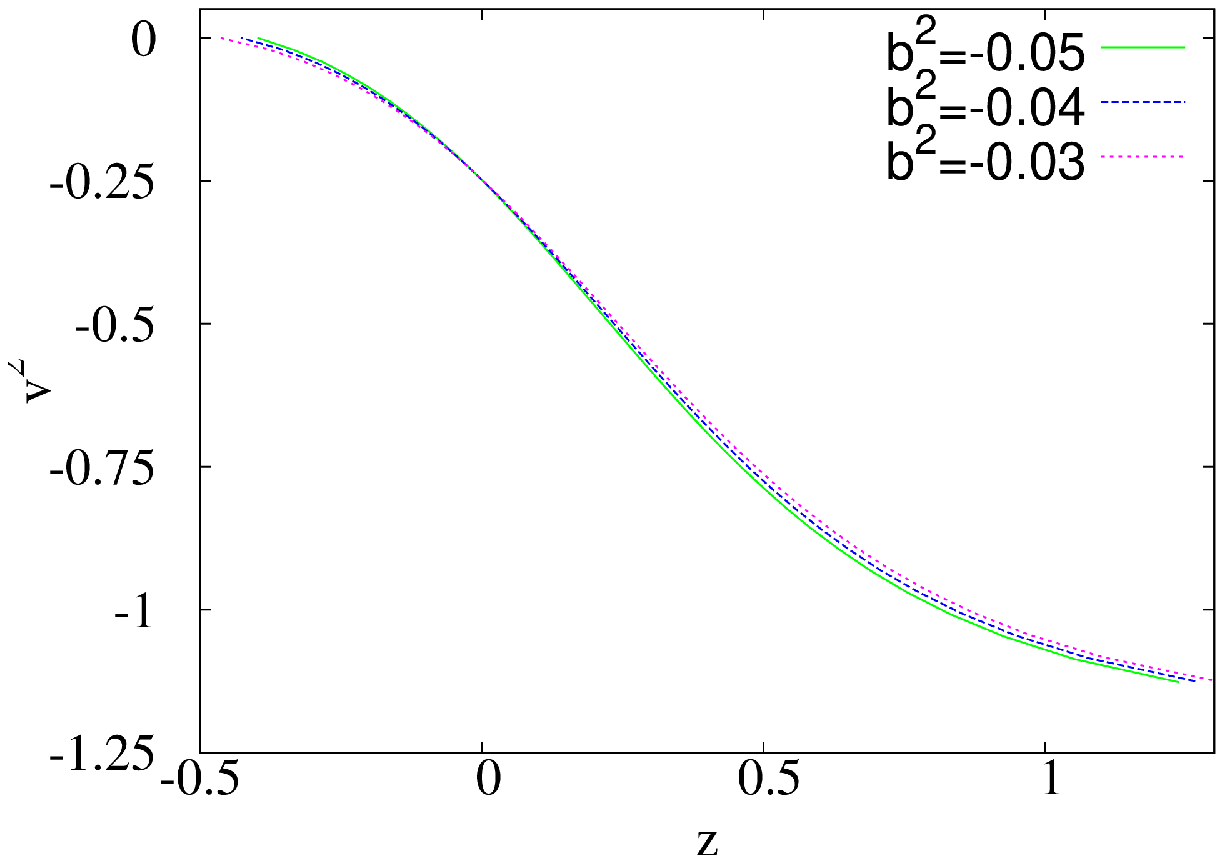}
 \caption{$v^{2}$ vs. $z$ plot for  $\delta=2.2$ and different $b^2$}
\label{vsqz}
\end{figure}
Following \cite{sharifkk} we consider $Q=\Gamma \rho_{D}$ and denote the ratio of the energy densities with $r$ ($r=\frac{\rho_m} {\rho_{D}}$). $\Gamma$ being the decay rate this gives the decay of GCG into CDM. An we effective equation of state can be defined (see \cite{setarehol}) as:
\begin{equation}
\label{efeosdef}
\omega^{eff}_{D}=\omega_{D}+\frac{\Gamma}{4H} \; \; and \; \; \omega^{eff}_{m}=-\frac{1}{r} \frac{\Gamma}{4H}
\end{equation}
Then we obtain from the continuity equations:
\begin{equation}
\label{kkcontm2}
\dot{\rho_m}+4H\rho_m(1+\omega^{eff}_m)=0
\end{equation}
\begin{equation}
\label{kkcontl2}
\dot{\rho_{D}}+4H\rho_D(1+\omega^{eff}_{D})=0
\end{equation}
Friedman equation in flat KK universe can be rewritten in terms of Hubble parameter:
\begin{equation}
\label{frwkkh}
H^2=\frac{1}{6} \left(\rho_{D}+\rho_m \right),
\end{equation}
where we set $M_p^2=1$. Density parameters are defined as:
\begin{equation}
\label{dpdef}
\Omega_m=\frac{\rho_m}{\rho_{cr}},  \Omega_{D}=\frac{\rho_{D}}{\rho_{cr}},
\end{equation}
where $\rho_{cr}=6H^2$. Eq. (\ref{frwkkh}) can be written in terms of density parameters as:
\begin{equation}
\label{frwdp}
\Omega_m+\Omega_{D}=1
\end{equation}
Using eqs. (\ref{dpdef}) and (\ref{frwdp}) we obtain:
\begin{equation}
\label{rdp}
r=\frac{1-\Omega_{D}}{\Omega_{D}}
\end{equation}
Now GCG (\ref{gcgeos}) is considered as a candidate of interacting dark energy model. In KK cosmology the energy density of GCG is given by:
\begin{equation}
\label{rhoa}
\rho_{D}=\left[ A+\frac{B}{a^{4(1+\alpha)}} \right]^\frac{1}{1+\alpha}
\end{equation}
The EOS parameters are  given by:
\begin{equation}
\label{oml}
\omega_{D}=\frac{p_{D}}{\rho_{D}}=-\frac{A}{\rho^{\alpha+1}_{D}}=-\frac{A}{A+\frac{B}{a^{4(1+\alpha)}}},
\end{equation}
\begin{equation}
\label{omlef}
\omega^{eff}_{D}=-\frac{A}{A+\frac{B}{a^{4(1+\alpha)}}}+\frac{\Gamma}{4H}.
\end{equation}
We consider a Tsallis holographic correspondence for GCG in KK cosmology following the prescription in \cite{thdemain,sharifkk} . As shown in \cite{sharifkk} the energy density for flat KK universe :
\begin{equation}
\label{rhohol}
\rho_{D}=B'L^{2\delta-4}
\end{equation}
The dark energy generated acceleration is a late time phenomenon. By then the additional dimension of KK theory would have shrunk enough and the expression in eq. (\ref{rhohol}) can be used for holographic correspondence. Effects of extra dimension are only manifested through the field equations. Infrared cutoff of the universe $L$ in the flat KK universe is equal to the apparent horizon, which coincides with Hubble horizon \cite{cai}. So we can write following \cite{sharifkk}:
\begin{equation}
\label{cutoff}
r_a=\frac{1}{H}=r_H=L
\end{equation}
The decay rate (\cite{wang}) is given by  ,:
\begin{equation}
\label{decay}
\Gamma=4b^2(1+r)H
\end{equation}
Making use of eq. (\ref{efeosdef}), (\ref{kkcontl2}), (\ref{rhohol}) and(\ref{decay}) on obtains: (\ref{omlef}) we obtain:
\begin{equation}
\label{omleffin}
\omega^{eff}_{D}=\frac{(1-\delta)-(2-\delta)b^2}{1-(2-\delta)\Omega_{D}}
\end{equation}
The correspondence between THDE and GCG  is drawn from eq. (\ref{rhoa}) and eq. (\ref{rhohol}):
\begin{equation}
\label{A1}
\left[A+\frac{B}{a^{4(1+\alpha)}}\right]^{\frac{1}{1+\alpha}}=B'H^{-2\delta+4}.
\end{equation}
$A$ is found from   eq. (\ref{rhohol}) and eq.  (\ref{omlef}):
\begin{equation}
\label{afin}
A=\frac{b^2-(1-\delta)\Omega_{D}}{\Omega_{D}\left[1-(2-\delta)\Omega_{D}\right]}\left(B'H^{-2\delta+4}\right)^{1+\alpha}
\end{equation}
Similarly, an expression for $B$ is found using eq. (\ref{afin}) in eq. (\ref{rhoa}),  and comparing it with eq. (\ref{rhohol}):
\begin{equation}
\label{bfin}
B =\frac{-(2-\delta)\Omega_{D}^2+(2-\delta)\Omega_{D}-b^2}{\Omega_{D}\left[1-(2-\delta)\Omega_{D}\right]}\left(B'H^{-2\delta+4}a^4\right)^{1+\alpha}
\end{equation}
Evolution of universe is summarized in fig. (\ref{omz}), (\ref{omeffz}) and (\ref{qz}), where the evolution of EOS parameter ($\omega^{eff}_D$), the density parameter $\Omega_D$ and deceleration parameter ($q$) are plotted with redshift ($z$). Present phase of acceleration is clearly incorporated in the model as seen form fig. (\ref{qz}).
\section{Squared Speed Of Sound in Chaplygin Gas and Classical Stability of the Model}
\label{sec:sqv}
\begin{figure}
\centering 
\includegraphics[scale=0.5]{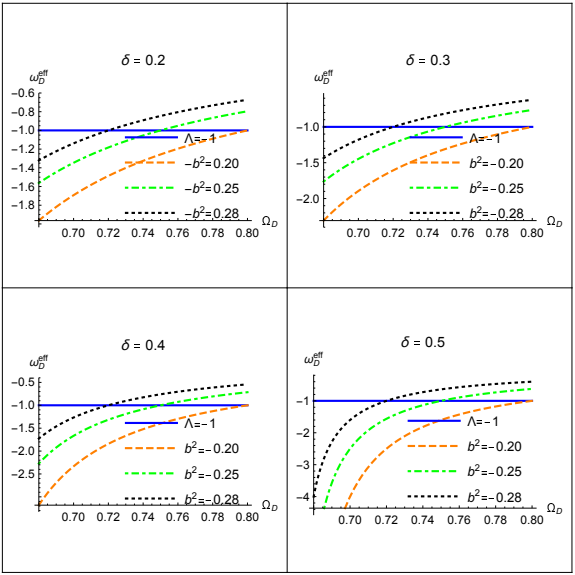}
\caption{Plot of $\omega_{D}$ vs. $\Omega_{D}$ for different values of $b^2$ (negative) and $\delta$}
\label{om:1}
\end{figure}
\begin{figure}
\centering
\includegraphics[scale=0.5]{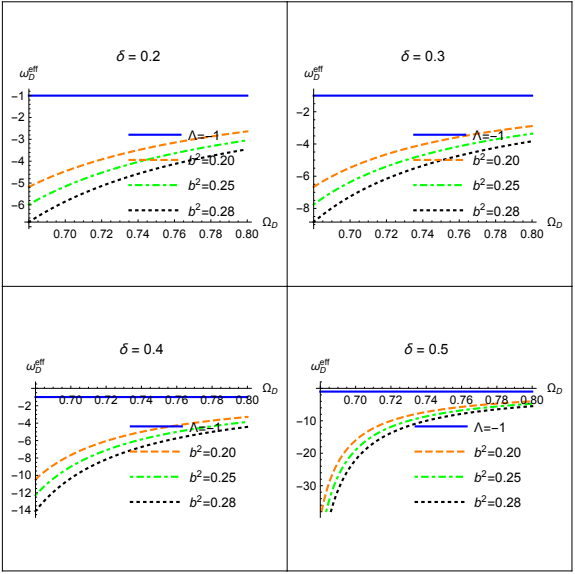}
\caption{Plot of $\omega_{D}$ vs. $\Omega_{D}$ for different values of $b^2$ (positive) and $\delta$}
\label{om:2}
\end{figure}
Squared adiabatic sound speed determines the  classical stability of the GCG model. The model is stable as long as the squared speed remains positive. Squared adiabatic sound speed is defined as:
\begin{equation}
\label{sqgcg}
v_{D}^2=\frac{dp_{D}}{d\rho_{D}}.
\end{equation}
The model is stable if $v_{D}^2 > 0$ \cite{setarehol}.  For our model this leads to:
\begin{equation}
\label{sqgcgchol}
v_{D}^2=\frac{dp_{D}}{d\rho_{D}}=\frac{\dot{p}}{\dot{\rho}}.
\end{equation}
$\dot{p}$ is given by:
\begin{equation}
\label{dotp}
\dot{p}=\dot{\omega}^{eff}_{D}\rho_{D}+\omega^{eff}_{D}\dot{\rho}_{D}
\end{equation}
where over dot is differentiation with respect to time. Hence,
\begin{equation}
\label{sqv}
v_{D}^2=\omega^{eff}_{D}+\dot{\omega}^{eff}_{D} \frac{\rho_{D}}{\dot{\rho}_{D}}.
\end{equation}
Using eq. (\ref{omleffin}) one obtains:
\begin{equation}
\label{dotom}
\dot{\omega}^{eff}_{D}=\frac{(2-\delta)\left[(1-\delta)-(2-\delta)b^2\right]}{\left[(1-(2-\delta)\Omega_{D})\right]^2} \dot{\Omega}_{D},
\end{equation}
where $\dot{\Omega}_{D}$ is fixed from  eqs. (\ref{dpdef}) and  (\ref{rhohol}). Using $L=\frac{1}{H}$, we get:
\begin{equation}
\label{dotom2}
\dot{\Omega}_{D}=\frac{1}{3}B'(1-\delta)H^{-2\delta +1}\dot{H}
\end{equation}
Substituting the values for $\dot{\omega}^{eff}_{D}$ and $\dot{\Omega}_{D}$ in eq. (\ref{sqv}), one obtains:
\begin{equation}
\label{sqvfin}
v_{D}^2=\omega^{eff}_{D}+\frac{(1-\delta)\left[(1-\delta)-(2-\delta)b^2\right]\Omega_{D}}{\left[1-(2-\delta)\Omega_{D}\right]^2}
\end{equation}
Eq. (\ref{sqvfin}) can be simplified by using eq. (\ref{omleffin}) to give:
\begin{equation}
\label{sqvfin2}
v_{D}^2=\omega^{eff}_{D}\left(\frac{1-\Omega_{D}}{1-(2-\delta)\Omega_{D}}\right)
\end{equation}
Equation (\ref{sqvfin2}) puts an upper bound on $\delta$ from stability consideration of the present model. Since $\omega_{D}^{eff}$ represents dark energy, $\omega_{D}^{eff}<0$. For the stability of our model  $v_{D}^2>0$ needs to be satisfied in the present epoch. Presently accepted value of $\Omega_{D}$ is $0.73$ and $1-\Omega_{D}>0$. For the stability of the present model $1-(2-\delta)\Omega_{D}<0$ is required. This leads to $\delta<0.63$. The stability is not much sensitive to the coupling parameter $b^2$ and remains stable in the present epoch as long as $\delta$ value remains within the discussed range.  However, the model becomes unstable beyond $\Omega_{D}=0.90$  irrespective of the value of $b^2$. Earlier we considered a $\delta$  value of $2.2$ which does not give a stable configuration throughout the evolutionary phases.  This is shown in fig (\ref{vsqz}). A different choice for IR cutoff or different kind of interaction may solve this problem. A stable configuration in the present epoch can be found considering $\delta<0.63$. This choice too can give acceptable values of $\omega^{eff}_D$ considering presently accepted value of $\Omega_D \approx 0.73$. This is shown in fig. (\ref{om:1}) and fig. (\ref{om:2}). However, it is noted that  in the present epoch ($\Omega_D \approx = 0.73$) the dark energy equation of state parameter has a value close to $-1$ ($\omega_D \approx -1$) as suggested by most of the observations. From fig. (\ref{om:1}) and fig. (\ref{om:2}) we see that for $b^2<0$, $\omega_D=-1$ is achieved for different $\Omega_D$ values. For lower values of $b^2$, $\omega_D=-1$ is obtained when $\Omega_D$ is close to $0.70$. For relatively higher values of of $b^2$, $\omega_D=-1$ is obtained when $\Omega_D$ is close to $0.8$. However, for $b^2>0$, $\omega_D<0$ irrespective of $\delta$ value.

\begin{figure}
\centering
 \includegraphics[scale=0.4]{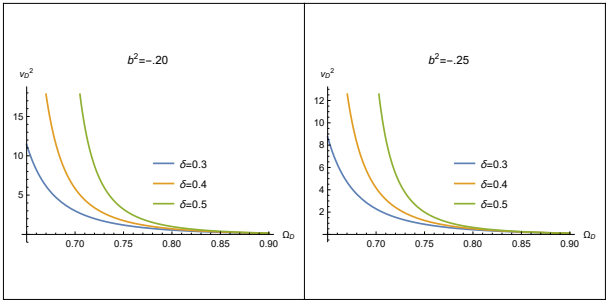}
 \caption{$v_{D}^{2}$ vs. $\Omega_{D}$ plot for different $\delta$ and $b^2$}
\label{vsq:1}
\end{figure}

\section{Discussion}
\label{disc}

THDE model with interacting Generalized Chaplygin Gas is considered here in KK Cosmology framework. It is seen from that the present model can successfully incorporate late time acceleration of universe (fig. (\ref{qz})). The model also gives acceptable value of EOS parameter (fig. (\ref{omeffz})) and dark energy density (fig. (\ref{omz})) for the present epoch ($z=0$). A few features of interest are noted: ($a$) For $b^2<0$ the energy density ceases to be phantom in nature during present epoch (fig. (\ref{om:1})). However it remains phantom for $b^2>0$ (fig. (\ref{om:2})) irrespective of choice of $\delta$ (although this effect is not completely insensitive to $\delta$). This is in contrast with \cite{ushol} where an usual HDE setup is considered. There phantom crossing appears for $b^2>0$. ($b$) The decay rate (the interaction term) may be computed at the present epoch from eq. (\ref{omleffin}) for a given $\omega^{eff}_{D}$. For example, as most of the observations suggest,  $\omega^{eff}_{D}=-1$ and $\Omega_{D}=0.73$ at the present epoch gives $b^2 \approx - 0.27$ for many different choices of $\delta$. ($c$) The HDE correspondence of interacting GCG model is unstable in $4D$ (\cite{setarehol}). It was later shown in \cite{ushol} that this correspondence become stable in the present epoch in KK. THDE correspondence of interacting GCG is also found stable in the present epoch (see section \ref{sec:sqv}). In particular, fig. (\ref{vsq:1}) shows that the model is stable for presently accepted value of $\Omega_D$. It is an interesting feature as the stability of THDE is not achieved with non interacting THDE models in $4D$ (see Ref. (\cite{thdemain})). For $\delta>0.63$ however, the model is not stable for any value of $b^2$. This issue may be addressed considering a different IR cutoff or a different kind of interaction or both.

\bibliographystyle{spr-mp-nameyear-cnd}
\bibliography{thderef}

\end{document}